\documentstyle[11pt,moriond,epsfig]{article}

\def\Journal#1#2#3#4{{#1} {\bf #2}, #3 (#4)} 
 
\def\NPB{{\em Nucl. Phys.} B} 
\def\PLB{{\em Phys. Lett.}  B} 
\def\PRL{\em Phys. Rev. Lett.} 
\def\PRD{{\em Phys. Rev.} D} 
\def\ZPC{{\em Z. Phys.} C} 
\def\EPJ{{\em Eur. Phys. J.} C} 
 \def\NPBps{{\em Nucl. Phys.} B (Proc. Suppl.)} 
\def\JHEP{\em J. High Energy Phys.}

\def\be{\begin{equation}} 
\def\ee{\end{equation}} 
\def\bea{\begin{eqnarray}} 
\def\eea{\end{eqnarray}} 
 
\begin{document} 
\begin{flushright}
{UB-ECM-PF-04/12}\\
{UG-FT-163/04}\\
{CAFPE-33/04}\\
\end{flushright}
\title{CP Violation in Kaons: $\varepsilon_K '/\varepsilon_K$ vs
  $K\rightarrow 3 \pi$} 
 
\author{I. Scimemi$^a$, E. G\'amiz$^b$, J. Prades$^c$ }

\address{$^a$ Facultat de F\'{\i}sica-ECM, Universitat de Barcelona\\ 
Av. Diagonal 647, E-08028 Barcelona, Spain\\ 
%
$^b$
Department of Physics $\&$ Astronomy,  University of
  Glasgow\\ Glasgow G12 8QQ, United Kingdom\\ 
$^c$ CAFPE and  
Departamento de F\'\i sica Te\'orica y del Cosmos, 
Universidad de Granada,\\ Campus de Fuente Nueva, E-18002 Granada, Spain}
 
\maketitle\abstracts{ The measurement 
of  $\varepsilon_K ' /\varepsilon_K$ has
  left open  several theoretical problems, the most important of which 
is an accurate calculation of the relevant hadronic matrix elements.
 Here we study the   possibility
of   relating $\varepsilon_K '$ to CP violation in 
charged $K\rightarrow 3 \pi$  which can provide information on the
reliability of such  calculations.
We discuss the r\^ole of final state interactions and the main sources
  of uncertainties.
We find that the measurement of $\Delta g_C$ in $K\rightarrow 3 \pi$ is a
 crucial consistency test  for the Standard Model calculations 
of $\varepsilon_K '$. 
Valuable information can be deduced also from other observables.} 
 
\section{Status of $ \varepsilon_K '$ and its theoretical calculations}
The experimental measurement of $ \varepsilon_K ' /\varepsilon_K$  by
KTeV~\cite{ktev:03} and by NA48~\cite{na48:03}  has proved  to be a
fundamental test for the  understanding of CP violation in the Standard
Model (SM).  The world average of the experimental 
results~\cite{ktev:03,na48:03,na31,e731} is
\bea
{\rm Re} \, 
\left(\frac{\varepsilon_K '}{\varepsilon_K}
\right) =(16.7\pm 1.6)\cdot 10^{-4} .
\eea

The theoretical prediction for this observable has  a long history 
and shown  to be rather difficult.
One needs several steps  in order to arrive to a final prediction.
\subsection{ Short-distance contributions}
The flavor changing phenomenon is born at a  scale of the
order of the gauge bosons mass, ${\cal O}(100 \ {\rm GeV})$,  which is
much bigger than the Kaon mass. Due to this difference
the gluonic corrections are amplified  by large logarithms.
These gluonic corrections are summed up using the
Operator Product Expansion (OPE) and renormalization group equations.
The running of the corresponding Wilson coefficients has been 
performed up to two
loops and is a well established result~\cite{buras1,ciuc1}.
The Lagrangian for $\Delta S=1$ processes  in the three flavor
theory is
\bea\label{eq:Leff}
 {\cal L}_{\mathrm eff}^{\Delta S=1}= \widetilde C \,  \sum_{i=1}^{10}
 C_i(\nu) \; Q_i (\nu)\ ; \; \quad \quad \widetilde
C= - \frac{G_F}{\sqrt{2}}
 V_{ud}^{\phantom{*}}\,V^*_{us}\ .
 \label{eq:lag}
\eea
The renormalization scale $\nu$  separates  the short-distance part
contained in the Wilson coefficients $C_i(\nu)$ and the long-distance
part encoded in the operators $Q_i(\nu)$. The final result  is
independent on this scale. The operators which are found to give the
main contributions to direct CP violation
 are the hadronic penguin $Q_6(\nu)$ and the electroweak penguin 
$Q_8(\nu)$.
\subsection{Long-distance}
The left   and most difficult 
problem is the  calculation of the  relevant hadronic matrix elements.
On the other hand Chiral Perturbation Theory (CHPT) 
is the effective field theory of the SM~\cite{GL:85,EC:95,PGR:86},
 which  describes the interaction of the low-lying
pseudo-Goldston bosons and  external sources.
At lowest order in the chiral expansion, $e^2 p^0$ and $e^0 p^2$, 
the effective realization of the Lagrangian in (\ref{eq:lag}) is 
\bea
\label{deltaS1}
{\cal L}^{(2)}_{|\Delta S|=1}&=&
\frac{3}{5}\,  \widetilde{C}
\,F_0^6 \, e^2 \, G_E \, \mbox{tr} \left( \Delta_{32} u^\dagger Q u\right)
+ \frac{3}{5}\, \widetilde{C} 
\, F_0^4 \left[ G_8 \, \mbox{tr} \left( \Delta_{32} u_\mu u^\mu \right)
+ G_8' \mbox{tr} \left( \Delta_{32} \chi_+ \right) \right.
\nonumber \\  &+& \left. 
G_{27} \, t^{ij,kl} \, \mbox{tr} \left( \Delta_{ij} u_\mu \right) \,
\mbox{tr} \left(\Delta_{kl} u^\mu\right) \right] + {\rm h.c.}
\eea
with $F_0$ the chiral limit value of the pion decay
constant $f_\pi= (92.4 \pm 0.4)$ MeV --see ref.~\cite{GPS:03} 
for definitions.
The real parts of the couplings $G_8$  and $G_{27}$
can be measured from the
CP conserving  observables  while their imaginary parts
are responsible for CP violation.
The theoretical determination of these  couplings is the 
challenge. The main couplings which come in the estimation of
$\varepsilon_K'$ are Im $G_8$ and Im $(e^2 G_E)$.
 The calculation of the $K \to \pi \pi$ amplitudes has been  done at NLO 
in CHPT \cite{BPP:98,PP:00,PPS:01} including isospin breaking
\cite{CENP:03}. 
In $\varepsilon_K'$  there  is  a cancellation  
between the contributions proportional to these two couplings
 which is very much reduced when all the relevant effects 
are taken into account.

\subsection{The unknown couplings}

At large $N_c$, all  the contributions 
to  Im $G_8$ and  Im $(e^2 G_E)$  are 
factorizable and the  scheme dependences are not  under control.
The unfactorizable topologies are not included at this order and they
bring in unrelated dynamics with its new scale and scheme
dependences, so that it is difficult to  give an uncertainty
to the large $N_c$ result for  Im $G_8$ and  Im $(e^2 G_E)$.

There are also calculations which take into account NLO large-Nc
corrections and also lattice computations.
The most recent results are  summarized  in Fig. 1.
Here the horizontal band 
represents the  region allowed by  the experimental value of
$\varepsilon_K'$. 
The rectangle on  the right is  the result from ~\cite{BP00,BGP01}. 
The rectangle on the left is the result from ~\cite{KPR01,HPR03}. 
The vertical band shows the lattice findings on
Im $(e^2\ G_E)$~\cite{domainwall,wilson}.
 The leading order large-$N_c$ result is marked with a small circle.

\begin{figure}[htbp]
\centering\begin{minipage}[t]{1cm}\vskip-4.8cm 
$\frac{{\rm Im} G_8}{{\rm Im} \tau}$\end{minipage}
\includegraphics[{width=6.5cm,height=6cm}]{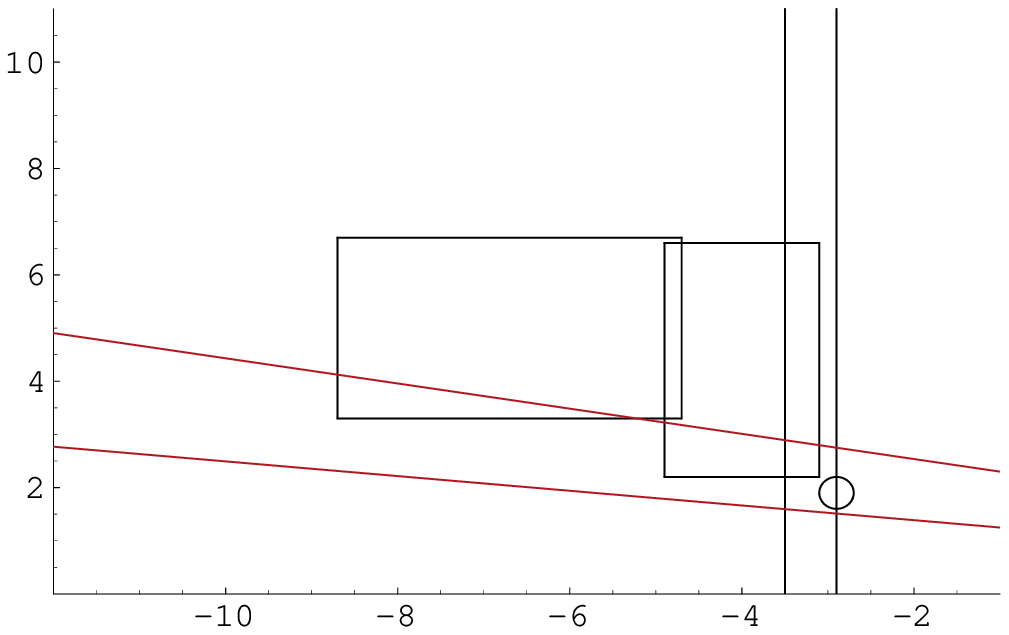}
\begin{minipage}[t]{1cm}\vskip-1.2cm $\frac{e^2{\rm Im} G_E}{{\rm Im}
    \tau}$\end{minipage}
\label{fig:status}
\caption{ Status of the most recent determination of the chiral weak
  couplings confronting $\varepsilon'_K$.
The  horizontal band is the allowed by 
the experimental result on $\varepsilon'_K$.
The theoretical results discussed in section 1  are depicted.}
\end{figure}

\section{CP violation observables in charged $K\rightarrow 3\pi$}

CP violating observables  in $K\rightarrow 3 \pi$
are  both experimentally  and theoretically very promising
and have attracted  a lot of recent effort (see ref.~\cite{GPS:03} for a complete discussion of other works~\cite{shabalin,DIPP94,DIP91,belkov}).

In \cite{GPS:03}, we  discussed CP-violating asymmetries
in the decay of the charged Kaon into three pions;
namely, asymmetries  in the slope $g$ defined as
\bea \label{gdefinition}
\frac{\left|A_{K^+\to 3 \pi}(s_1,s_2,s_3)\right|^2}
{\left|A_{K^+\to 3 \pi}(s_0,s_0,s_0) \right|^2}=
1+ g \, y + h \, y^2 + k \, x^2 + {\cal O}( y  x^2,y^3) \, 
\eea
 and some asymmetries in the integrated $K^+ \to  3 \pi$ decay rates. 
Above,  we used the Dalitz variables
$ x\equiv \frac{s_1-s_2}{m_{\pi^+}^2}$
and 
$y \equiv \frac{s_3-s_0}{m_{\pi^+}^2} $
with 
$s_i\equiv(k-p_i)^2$, $3s_0\equiv m_K^2 + m_{\pi^{(1)}}^2+ 
m_{\pi^{(2)}}^2+m_{\pi^{(3)}}^2$. 
The CP-violating asymmetries in the slope $g$ 
are defined as 
\bea
\label{defDeltag}
\Delta g_C \equiv
\frac{g[K^+ \to \pi^+ \pi^+ \pi^-]-g[K^-\to\pi^-\pi^-\pi^+]}
{g[K^+ \to \pi^+ \pi^+ \pi^-]+g[K^-\to\pi^-\pi^-\pi^+]}
\nonumber \\ {\rm and} \hspace*{0.5cm} 
 \Delta g_N \equiv 
\frac{g[K^+ \to \pi^0 \pi^0 \pi^+]-g[K^-\to\pi^0\pi^0\pi^-]}
{g[K^+ \to \pi^0 \pi^0 \pi^+]+g[K^-\to\pi^0\pi^0\pi^-]} \, .
\eea
The CP-violating asymmetries in the decay rates, $\Delta \Gamma_{C(N)}$, are defined analogously.

 Recently, two experiments, namely, NA48 at CERN and KLOE at Frascati,
have announced the possibility of measuring
the asymmetries $\Delta g_C$  and 
$\Delta g_N$ with a sensitivity of the order of $10^{-4}$,
i.e., two orders of magnitude better than at present \cite{ISTRA+}, 
 see for instance \cite{NA48} and \cite{KLOE}.
It is therefore mandatory to have 
these predictions  at  NLO in CHPT as they are provided in 
~\cite{GPS:03}. 

\subsection{CP-Violating Predictions at Leading Order} 
\label{LOresults}
The slopes $g$ and decay rates $\Gamma$ start at  order $p^2$ in CHPT.
We have used the LO result to make some checks.
We have checked that
the effect of  Re $(e^2 G_E)$  is very small
also for the $\Delta g$ and $\Delta \Gamma$ asymmetries. 
We have also checked that the asymmetries 
$\Delta g_{C (N)}$ are very  poorly sensitive to Im $(e^2 G_E)$.
This fact makes  an accurate enough measurement of these asymmetries 
 very interesting to check if  
Im $G_8$ can be as large as predicted in \cite{BP00,HPR03,HKPS00}.
It also makes these CP-violating asymmetries
complementary to the direct CP-violating parameter $\varepsilon'_K$.
The same poor dependence on ${\rm Im}  (e^2 G_E)$ is observed in 
the LO  result for the total widths asymmetries~\cite{GPS:03}.
However, in order to make a  more precise analysis, it is convenient 
to go at NLO in  CHPT.

\subsection{CP-Violating Predictions at Next-to-Leading Order} 
At NLO one needs the real parts of the amplitudes at order $p^4$ and
the FSI at order $p^6$.
 The real part of the amplitudes for the octet and 27-plet 
components were recently
 computed   in \cite{BDP03} and checked by us in ~\cite{GPS:03}. 
The electroweak  part and  the relevant FSI were
 computed  in \cite{GPS:03}. We refer to these papers for the details.

To describe $K\to 3 \pi$ at NLO, 
in addition to Re $G_8$, $G_{27}$, Re $(e^2 G_E)$, Im $G_8$ and
Im $(e^2 G_E)$, we also need several other ingredients.
Namely, for the real part we need the chiral logs and the counterterms. 
The relevant counterterm combinations 
were called $\widetilde K_i$ in \cite{BDP03}. 
 The real part of the counterterms,
 Re $\widetilde K_i$, can be obtained from 
the fit  of the $K \to 3 \pi$ 
CP-conserving decays  to data done in \cite{BDP03}. 

The imaginary part 
of the order $p^4$ counterterms, Im $\widetilde K_i$,
is much more problematic.  They cannot be obtained from data and 
there is no available calculation for them at NLO in $1/N_c$ .
 A direct calculation of them  at NLO in $1/N_c$ 
can be done using the appropriate hadronic Green functions
--a scheme to get them 
has been setted up recently in \cite{BGLP03}.

One can use several approaches to get the order of magnitude
and/or the signs of ${\rm Im} \widetilde K_i$.
We will follow here a more naive approach
that will be enough for our purpose of estimating the effect of the 
unknown counterterms.
 We can assume   that the ratio of the real to the imaginary parts
is  dominated by the same strong dynamics at LO and NLO 
in CHPT, therefore
\bea
\label{assum1}
\frac{{\rm Im}  \widetilde K_i}{{\rm Re} \widetilde K_i}
\simeq \frac{{\rm Im}  G_8}{{\rm Re} G_8} 
\simeq \frac{{\rm Im}  G_8'}{ {\rm Re} G_8'} \simeq (0.9 \pm 0.3) 
\, {\rm Im}  \tau  \, .
\eea
In particular, we set to zero those Im $\widetilde K_i$
whose corresponding Re  $\widetilde K_i$ are 
set also to zero in the fit to CP-conserving amplitudes
done in \cite{BDP03}.
Of course, the relation above can only be applied to those 
$\widetilde K_i$ couplings with non-vanishing imaginary part.
Octet dominance to order $p^4$ is a further
assumption implicit in (\ref{assum1}).
The second equality  in (\ref{assum1})
 is well satisfied by the model calculation in \cite{BP00}, which is 
the only full calculation at NLO in $1/N_c$ at present. 

The values of Im $\widetilde K_i$ obtained using (\ref{assum1}) 
 will allow us to check the counterterm dependence
of the CP-violating asymmetries. They will also provide us  a good 
estimate of the counterterm contribution to the CP-violating asymmetries
that we are studying. 

The final results for the slope asymmetries are 
\bea
\label{eq:gNLOeff}
\frac{
\Delta g_{C}}{10^{-2}}
\!\!\!&\simeq&\!\!\!\! \left \lbrack (0.66\pm 0.13) \, 
{\rm Im}  G_8 +(4.3\pm 1.6)\, {\rm Im}  \widetilde K_2 -(18.1\pm 2.2) \, {\rm Im}  \widetilde K_3 
 -  (0.07\pm 0.02)\,{\rm Im}  (e^2 G_E) \right \rbrack 
 ,
\nonumber \\ 
\frac{\Delta g_{N}}{10^{-2}}
\!\!\! &\simeq&\!\!\!\!  - \left \lbrack (0.04\pm 0.08) \, {\rm Im}  G_8 
+ (3.7\pm 1.1)\, {\rm Im}  \widetilde K_2  
+ (26.3\pm 3.6)\, {\rm Im}  \widetilde K_3 \right.
\nonumber\\
\!\!\! &&\!\!\!\!\left.
+ (0.05\pm 0.02) \, {\rm Im}  (e^2 G_E) 
\right \rbrack 
\, . 
\eea
Similar numerical formulas can be deduced also for the width asymmetry
and can be found in ref.~\cite{GPS:03}. From these formulas 
one deduces that $\Delta g_C$  is  a quite clean
observable to measure Im $G_8$ and so cross-check the result of
$\varepsilon_K'$.
Using the inputs of ref.~\cite{BGP01} for ${\rm Im}G_8$ and ${\rm Im} 
(e^2\  G_E)$ and the estimation of the 
counterterms contributions of ref.~\cite{GPS:03} one finds
\bea
\Delta g_C=-(2.4\pm 1.2)\cdot 10^{-5}\ ;
\quad\quad \Delta g_N=(1.1\pm 0.7)\cdot 10^{-5} \ .
\eea
\noindent
\begin{figure}\begin{minipage}[t]{.7cm} \vskip-7cm$\frac{{\rm Im}
      G_8}{{\rm Im} \tau}$
\end{minipage}
\begin{minipage}[b]{.40\linewidth}
\centering\epsfig{figure=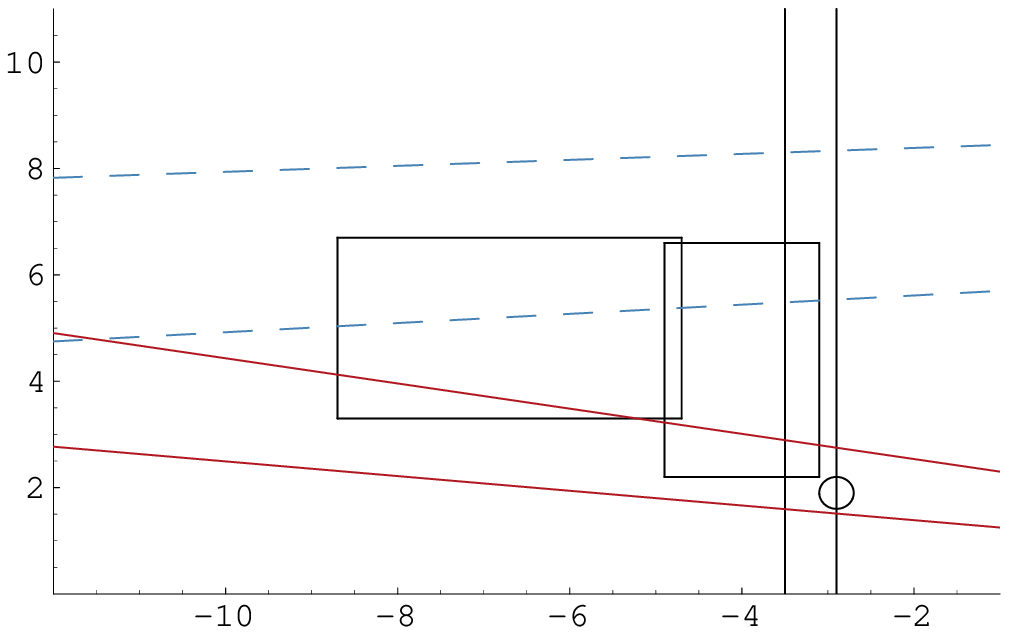,width=\linewidth}
\caption{$\Delta g_C\sim -3.5\cdot
 10^{-5}$ (band between blue dashed lines) confronting
the experimental value of  
$\varepsilon_K'$ (band between red solid lines) and its theoretical estimates 
(see text).}\label{fig:gchigh}
\end{minipage}\hfill
\begin{minipage}[t]{.7cm} \vskip-7cm$\frac{{\rm Im}
      G_8}{{\rm Im} \tau}$
\end{minipage}
\begin{minipage}[b]{.40\linewidth}
\centering\epsfig{figure=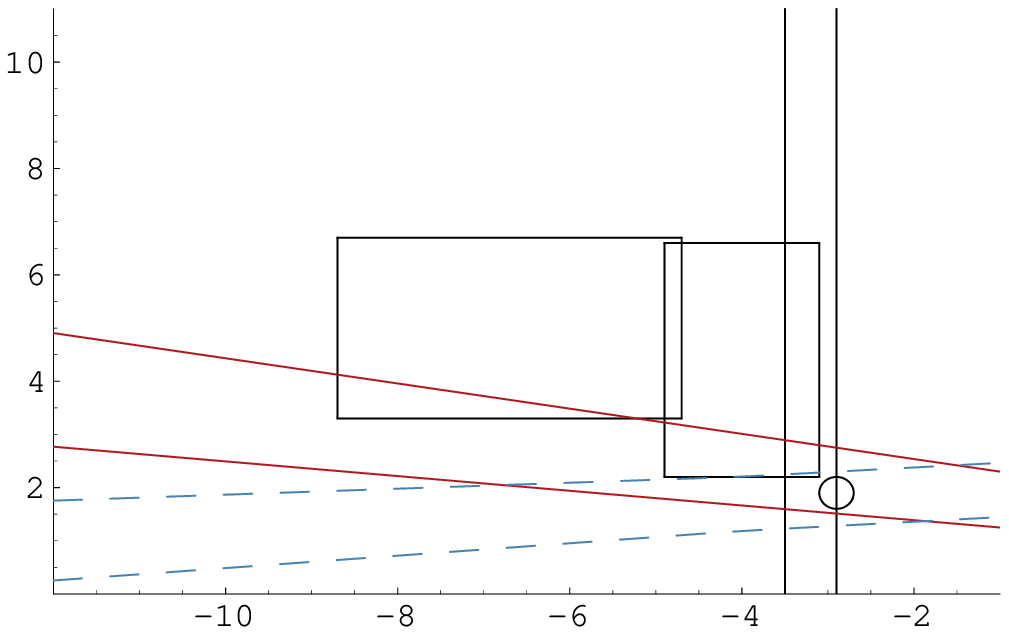,width=\linewidth}
\caption{$\Delta g_C\sim - 10^{-5}$ (band between blue dashed lines) confronting 
the experimental value of 
$\varepsilon_K'$ (band between red solid lines) and its theoretical estimates 
(see text).}\label{fig:figc}
\end{minipage}\hfill
\begin{minipage}[t]{.7cm}\vskip-3cm\hspace{-9.2cm}$\frac{{\rm Im}
     e^2 G_E}{{\rm Im} \tau}$ \hspace{6.3cm}$\frac{{\rm Im}
     e^2 G_E}{{\rm Im} \tau}$
\end{minipage}
\end{figure}
\section{Conclusions}
 The impact of the eventual measurement of $\Delta g_C$  is depicted in
 Fig.~\ref{fig:gchigh} and in Fig.~\ref{fig:figc}.
A value of $10^5\cdot \Delta g_C<-4$ would  be  a sign of new physics. 
For  $-2>10^5 \cdot \Delta g_C>-4$ the high values of ${\rm Im} G_8$
 would be confirmed, although the measure would not be in straight
 agreement with the measurement of $\varepsilon_K'$.
 A  value of $10^5\cdot \Delta g_C\sim -1 $ 
would  fit perfectly the SM value
 of $\varepsilon_K'$. An experimental precision of the
 order of $ (1\sim1.5)\cdot 10^{-5}$ would be required.

The asymmetry  $\Delta g_N$ and the asymmetries in the  
total widths would give also important informations  
on  the counterterms magnitude
 but not so much on Im $G_8$  and Im $(e^2 G_E)$.
An example is provided in Fig. 4. In this figure we 
use 
\bea
\label{assum2}
\frac{{\rm Im}  \widetilde K_2}{{\rm Re} \widetilde K_2}
\simeq \omega  \frac{{\rm Im}  \widetilde K_3}{{\rm Re} \widetilde K_3}
\simeq k \frac{{\rm Im}  G_8}{{\rm Re} G_8} 
\eea
 and we  plot  $\Delta g_N \cdot 10^5$ 
versus $k$ for several values of the parameter  $\omega$ .
 An upper limit on $\Delta g_N$ of $10^{-5}$ would already put a
 valuable constraint on $k$, namely $k<3$. 
The total width asymmetries are
 even  more difficult to predict~\cite{GPS:03}.

The measurement of all these charged $K \to 3 \pi$
CP asymmetries  would certainly be 
an experimental achievement   and  would provide
extremely  precious information 
for the understanding of the mechanism of CP-violation.
\section*{Acknowledgments}
This work has been supported in part by the
European Union RTN Network EURIDICE under Contract No.
HPRN-CT2002-00311. The work of I.S. 
has been supported in part by MEC (Spain) and FEDER
(European Union)  Grant No. FPA2001-03598. 
 E.G. is indebted to the European Union for a
Marie Curie Intra-European Fellowship.
The work of J.P. has been supported in by 
MEC (Spain) and FEDER (European Union) 
Grant No.  FPA2003-09298-C02-01
and by  Junta de Andaluc\'{\i}a Grant No. FQM-101.

\appendix
\section*{Misprints in  reference ~\cite{GPS:03}}
We report here some  misprints \footnote{We thank Hans Bijnens, 
Pierre Dhonte and Fredrik Persson for pointing them to us.} 
found in our original paper \cite{GPS:03}. 
The signs of the rhs of Eqs. B.12 and B.15 must be flipped except
for $A_c$. In Eq.  B.29 the counterterm $Z_8^r$ must be
proportional to $9(m_K^2-2 m_\pi^2)$ and not to $9(m_K^2-m_\pi^2)$.
\begin{figure}[htbp]
\centering\begin{minipage}[t]{1.5cm}\vskip-4.8cm $\Delta g_N\ 10^5$\end{minipage}
\includegraphics[{width=6.5cm,height=6cm}]{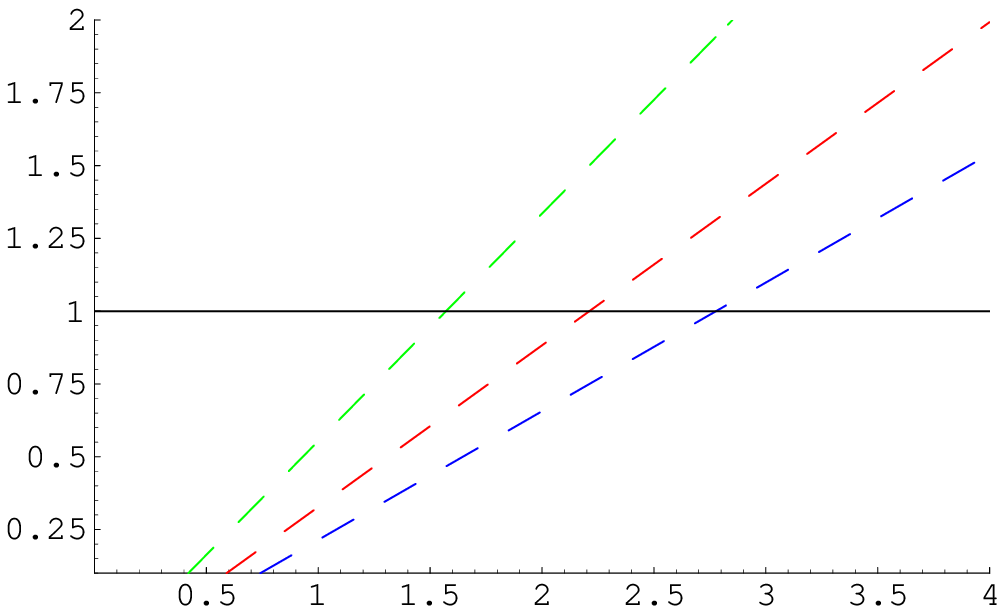}
\begin{minipage}[t]{1cm}\vskip-1cm $k$\end{minipage}
\label{fig:gnlim}
\caption{Constraints on the counterterms coming from 
an eventual upper limit of $\Delta g_N$ of $10^{-5}$ (horizontal line).
 The three dashed lines correspond respectively from left to right
 to $\omega=1/2,\ 1,\ 2$. }
\end{figure}

\end{document}